\documentclass[a4paper]{article}

\usepackage[english]{babel}
\usepackage[utf8x]{inputenc}
\usepackage[T1]{fontenc}
\usepackage{booktabs}
\usepackage{rotating}
\usepackage{float}
\usepackage{multirow}
\usepackage{comment}
\usepackage{makecell}
\usepackage[table]{xcolor}
\usepackage[flushleft]{threeparttable} 
\usepackage{environ}
\usepackage{microtype}

\usepackage[a4paper,top=3cm,bottom=2cm,left=3cm,right=3cm,marginparwidth=1.75cm]{geometry}

\usepackage{amsmath}
\usepackage{url} 
\usepackage{array}
\usepackage{tabulary}
\usepackage{tabularx,colortbl}
\usepackage{graphicx}
\usepackage[colorinlistoftodos]{todonotes}
\usepackage[colorlinks=true, allcolors=blue]{hyperref}
\usepackage{hhline}
\usepackage{xcolor}
\usepackage{authblk}

\newcommand\crule[3][black]{\textcolor{#1}{\rule{#2}{#3}}}

\newcommand\blfootnote[1]{%
  \begingroup
  \renewcommand\thefootnote{}\footnote{#1}%
  \addtocounter{footnote}{-1}%
  \endgroup
}

\begin{document}

\title{A growth adjusted price-earnings ratio}
\author{Graham Baird\thanks{Mathematical Institute, University of Oxford}, James Dodd\thanks{St. Cross College, University of Oxford}, Lawrence Middleton\thanks{Department of Statistics, University of Oxford}}
\date{\today}
\maketitle

\begin{abstract} \noindent The purpose of this paper is to introduce a new growth adjusted price-earnings measure (GA-$P/E$) and assess its efficacy as measure of value and predictor of future stock returns. Taking inspiration from the interpretation of the traditional price-earnings ratio as a period of time, the new measure computes the requisite payback period whilst accounting for earnings growth. Having derived the measure, we outline a number of its properties before conducting an extensive empirical study utilising a sorted portfolio methodology. We find that the returns of the low GA-$P/E$ stocks exceed those of the high GA-$P/E$ stocks, both in an absolute sense and also on a risk-adjusted basis. Furthermore, the returns from the low GA-$P/E$ porfolio was found to exceed those of the value portfolio arising from a $P/E$ sort on the same pool of stocks. Finally, the returns of our GA-$P/E$ sorted porfolios were subjected to analysis by conducting regressions against the standard Fama and French risk factors.
\end{abstract}

\section{Introduction}

\blfootnote{Research sponsored by The Witness Corporation (www.thewitnesscorporation.com)}

\noindent The classical strategy of value investing involves purchasing stocks which are deemed to be undervalued relative to their intrinsic value. In practice, when it comes to determining whether a given stock is undervalued or not, investors typically rely on a number of standard value metrics, for example,  a stock possessing a high book-to-market ($B/M$), or alternatively a low price-earnings ratio ($P/E$) would generally be seen as a `value stock'. At the other end of the spectrum are `growth stocks' which typically trade at high multiples of both book value (low $B/M$) and earnings (high $P/E$). Generally investors are willing to pay this higher price, relative to the current intrinsic value, as it  is anticipated that the company will experience significant growth in their earnings and future book value. Often growth stocks are associated with newly listed companies operating in rapidly expanding industries, such as technology and healthcare.

The value approach has long been employed by investors, with Benjamin Graham and David Dodd being amongst the earliest proponents of the style, espousing its merits at length in the seminal text Security Analyis \cite{graham}. This led to Graham being given the title ``the father of value investing'', and the pair have attracted many notable followers within the investment community. There has also been a great deal of debate within the academic literature regarding both the general efficacy of value investing and also in explaining the strategy's historical 
performance. Focussing on the price-earnings ratio in particular, significant works include Nicholson \cite{nicholson}, who presented the first study to systematically demonstrate a `$P/E$ effect', whereby stocks with lower price-earnings ratios provide greater subsequent returns than those with higher price-earnings ratios. The later work by Basu \cite{basu77}, offered more comprehensive evidence for the existence of such a $P/E$ effect, with the author further demonstrating that the outperformance persists even when returns are adjusted for risk. In the intervening period since these early articles there has been much discussion as to the veracity of the price-earnings effect. Some have suggested that the price-earnings ratio simply acts as a proxy for other factors \cite{ball78}, for example firm size \cite{reinganum81}, or that the observed effect occurs as a consequence of biases in the datasets utilised \cite{banzgreen86}, whilst other studies have continued to support the existence of the price-earnings effect \cite{jaffe89}.

It is now generally accepted that stocks with low price-earnings ratios offer higher returns than those with a high price-earnings ratio, a finding consistent across markets and time periods. Furthermore, this represents just one manifestation of a wider value effect, with similar results obtained across the class of value metrics \cite{fama98}. This being the case, the discussion has largely turned to providing an explanation for the additional returns offered by value stocks. One  explanation, advanced most notably by Eugene Fama and Kenneth French, is that value stocks tend to be associated with firms which have experienced a period of distress, and as such provide inherently riskier investment opportunities \cite{fama92}. In light of the higher risk, the additional return observed with value stocks simply represents the necessary compensation required by the market. This has led to the inclusion of a `value risk factor' in  a number of multi-factor extensions of the traditional capital asset pricing model (CAPM), for example Fama and French's own three \cite{fama93} and five \cite{fama15} factor models, and Carhart's four factor model \cite{carhart97}. Alternatively, some authors, for example De Bondt and Thaler \cite{thaler85,thaler87}, Chopra, Lakonishok and Ritter \cite{ritter92}, Lakonishok, Shleifer, and Vishny \cite{lsv94} and Haugen \cite{haugen95} argue that this additional return for value stocks arises because the market has a tendency to overreact to temporary short-term circumstances. In particular, undervaluing currently distressed stocks whilst overvaluing the stocks of recently successful firms (growth or glamour stocks). When the circumstances are resolved or the exceptional success cannot be sustained, these pricing errors are corrected and the distressed (value) stocks experience higher returns in comparison to growth stocks. 

As discussed, traditional value stocks are those which trade at a low price relative to their intrinsic value, a value which is determined largely with respect to the current circumstances of the firm. However, one might argue that in order to determine the true intrinsic value of a stock it is necessary to take account of the growth prospects of the underlying firm. If such prospects are taken into account within our calculations, then we might expect to find true value prospects across the traditional value/growth spectrum. It is the purpose of this paper to propose a new `value' measure which takes account of both present value and earnings growth. Having introduced the measure and examined its properties, we examine its efficacy as a measure and predictor of future returns through an extensive empirical study, covering stocks listed on the the NYSE, Nasdaq and AMEX over the period 1990 to 2015. However, we begin by reviewing the presently available price-earnings based measures and outlining some of their features.

\section{Existing price-earning measures and accounting for growth}

\subsection{The price-earnings ratio and its variants}
\noindent Of the range of value measures, the price-earnings ratio or $P/E$-ratio is perhaps the most widely discussed metric when it comes to stock valuation. Defined as the ratio of a company's stock price to (some measure of) its annual earnings per share (EPS), the price-earnings ratio may be viewed as the price investors must pay per unit of current earnings, or alternatively as the period required in order for a stock to accrue its current value in earnings, assuming that earnings remain constant. Whilst an apparently simple measure, multiple variants of the $P/E$-ratio exist, with these mostly varying in the earnings figure used within the calculation. 

When the earnings figure is derived from the previous fiscal year, or alternatively the previously reported four quarters, then it is common to refer to the ratio as the `trailing' $P/E$-ratio. The advantage of this variant is that it provides an objective measure of value. However, by nature it is a backward looking measure, which may make it less relevant in predicting future stock returns. 

Alternatively, it is common to use estimates of the next set of annual earnings, or else predictions of earnings over the following four quarters. Calculating the ratio using such figures provides us with the `forward' $P/E$-ratio. The requisite forward earnings figures are typically estimated  by averaging the predictions of a group of market analysts. By using the forward earnings the objective is to provide a measure that is more representative of the future prospects of the firm, however the downside is that we introduce a subjective element to our measure.

A further variant on the $P/E$-ratio is the cyclically adjusted price-earnings ratio (CAPE). Instead of using a single year's earnings in the calculation, the CAPE utilises the average of the previous ten years earnings, after adjusting for inflation. In averaging the earnings, the objective is to reduce the effect of earnings volatility over the course of the business cycle, which may prove beneficial especially when making longer term forecasts of future returns. The CAPE was popularised by Robert Shiller and John Campbell \cite{shiller88,campbell98}, who applied the CAPE on a stock index level, however the idea of averaging earnings over a number of years in order to stabilise the $P/E$ value goes back at least to Graham and Dodds \cite{graham}.

In theory the price-earnings ratio allows an investor to make value judgments between stocks, however caution must be exercised. Intuitively we might expect that investors should be willing to pay a higher multiple of current earnings 
only if they expect the stock to reliably produce higher earnings going forward. Such a hypothesis is supported by those studies which set out to derive a theoretically `fair' $P/E$ value, for example \cite{gordon62, malkiel70} and \cite{fairfield94}. However, the empirical evidence \cite{beaver78,penman96,thomas06} is mixed as to whether the observed discrepancies in $P/E$-ratios may be justified by growth. In any event, na\"{i}vely comparing vastly differing stocks based solely on price-earnings ratio is likely to provide the investor with an incomplete picture of their comparative value.

Having outlined a number of the variants of the standard price-earnings ratio, and its unsuitability for making value judgments between stocks with differing growth expectations, we now spend some time examining a number of models and metrics which attempt to incorporate growth expectations with a measure of value.

\subsection{PEG ratio and PEG payback period}
As discussed, investors are generally willing to accept a higher $P/E$-ratio when they believe a firm has credible prospects for strong earnings growth. Therefore, when assessing a stock, investors must weigh up the value it currently offers with their expectations for earnings growth. The price-earnings to growth ratio, or PEG ratio, attempts to quantify this trade-off between growth and current value, enabling a more ready comparison to be made between stocks. The PEG ratio is derived from the $P/E$-ratio in the following manner:

\begin{equation}\label{PEGratio}
\text{PEG Ratio}=\frac{P/E\text{ Ratio}}{\text{Annual EPS growth rate in \%}}.\\
\end{equation}
As with the $P/E$-ratio, there are a number of variants of the PEG ratio depending on the $P/E$-ratio used in the calculation and how the earnings growth rate is determined. If the trailing $P/E$-ratio is combined with a growth rate estimated from historical earnings, then it is common to refer to the resulting PEG as the \textit{trailing} PEG ratio. However, if the forward $P/E$-ratio is used alongside a forecast of future earnings growth, then we obtain the \textit{forward} PEG ratio. Once again the forward earnings and growth rate are obtained from surveys of professional analysts' forecasts.

Originally developed by Farina in \cite{farina69}, the PEG ratio was most notably lauded by Peter Lynch in his book \textit{One Up On Wall Street} \cite{lynch}. In this book, Lynch states that ``The $P/E$-ratio of any company that's fairly priced will equal its growth rate", implying that a fairly priced stock should have a PEG ratio of $1$. With this rule of thumb, stocks with a PEG ratio below $1$ are deemed to provide good value, whilst those with a PEG above $1$ are seen as overpriced. However, as opposed to the $P/E$-ratio which we may interpret as the price of future earnings or alternatively as a payback period, the PEG ratio is simply a heuristic measure and lacks any real intrinsic significance in terms of valuation \cite{easton04}. In the upcoming example featured in Table \ref{earngrow}, we shall demonstrate the shortcomings of the PEG ratio. The two stocks considered both have a PEG of $1$, however they are shown to provide unequal investment opportunities.

An alternative measure which attempts to account for earnings growth within stock valuation is the PEG payback period \cite{MotFoolPEGP,MorningStarPEGP}. The PEG payback period follows similar lines to the $P/E$-ratio in its interpretation as a time period, however, the PEG payback period accounts for earnings growth. Simply put, the PEG payback period is the minimum number of whole years a stock must be held in order for its accrued earnings to equal or exceed its initial price, assuming that the annual earnings grow at a constant rate $g$. However, it is somewhat naive in its approach, typically proceeding via a simplistic calculation. 

For example, let us consider a stock with price $P=10$ and earnings per share of $E=1$ over the past four quarters, hence the stock has a trailing $P/E$-ratio of $10$. If we assume that earnings per share grow at a constant $10\%$ per annum, then the table below (Table~\ref{earngrow}) sets out the future annual earnings and associated cumulative earnings for the stock. From this table we can see that after $7$ years the cumulative earnings of the stock surpass the initial stock price, hence the stock has a PEG payback period of $7$. In comparison, if we consider a stock with $P=20$ and $E=1$ with $20\%$ earnings growth then we can see that the initial price is only recouped after $9$ years, by which point the $P/E=10$ stock has earned $149\%$ of the initial price, in comparison to $125\%$ for the $P/E=20$ stock. As such, these two stocks with the same PEG ratio clearly do not offer the same value to the investor. However, if earnings continued to grow at the given rates then the $P/E=20$ stock would soon surpass the $P/E=10$ stock in terms of its return. This example also clearly demonstrates the imprecision of the PEG payback period as a metric of value. Examining the earnings schedule for the stock with $P=20$, after $8$ years we can see cumulative earnings have almost reached $20$. However, we require an additional year to surpass $P$ in full, by which point cumulative earnings exceed $P$ by almost $25\%$. As with the PEG ratio, the PEG payback period originates more from the popular investment literature, and to date has received little or no coverage in the academic literature.

\begin{table}[h]
\begin{center}
{\begin{tabular}{ccccc} \toprule
\rule{0pt}{10pt} & \multicolumn{2}{c}{$P/E=10$, with 10\% growth} & \multicolumn{2}{c}{$P/E=20$, with 20\% growth}  \\ \cmidrule{2-5}
\rule{0pt}{10pt} Year      & Earnings & Cumulative & Earnings & Cumulative  \\ \hline
\rule{-3pt}{10pt}
0         & 1.00     &            &  1.00    &             \\
1         & 1.10     & 1.10       &  1.20    & 1.20        \\
2         & 1.21     & 2.31       & 1.44     & 2.64        \\
3         & 1.33     & 3.64       & 1.73     & 4.37        \\
4         & 1.46     & 5.11       & 2.07     & 6.44        \\
5         & 1.61     & 6.72       & 2.49     & 8.93        \\
6         & 1.77     & 8.49       & 2.99     & 11.92       \\
7         & 1.95     & \textbf{10.44}      & 3.58     & 15.50       \\
8         & 2.14     & 12.58      & 4.30     & 19.80       \\
9         & 2.36     & 14.94      & 5.16     & \textbf{24.96}       \\
\bottomrule
\end{tabular}}
\end{center}
\caption{Future earnings for stocks with $P/E$'s of 10 and 20 and growth rates of 10\% and 20\% respectively.}
\label{earngrow}
\end{table}

\subsection{Modelling stock value as a function of earnings and growth}
The previous measures take a stocks price, the earnings per share and earnings growth rate as given and attempt to ascertain whether it represents a favourable investment opportunity. However, there exists another group of models which take the earnings per share and growth rate, amongst other variables, and look to derive a fair price for the underlying stock, having done so it is then possible to obtain a theoretically `correct' $P/E$-ratio for the stock. The first such model derives from the Gordon growth model \cite{gordon62}, which equates a stocks worth to the sum of all of its future dividend payments, discounted back to their present value. Letting $D$ denote the most recent dividend paid by the stock, the model assumes that the stock will pay a dividend in all future years, with a constant annual dividend growth rate of $g_D$. If $r(>g_D)$ represents the cost of equity capital for the firm, that being the rate at which we will discount future payments, then the model gives the following value for the stock: 
\[P=\sum_{n=1}^\infty\left(\frac{1+g_D}{1+r}\right)^{\hspace{-1mm}n}\hspace{-1mm}D=\frac{(1+g_D)D}{r-g_D}.\]
Assuming that the firm pays a constant proportion $\gamma$ of earnings as dividends each year, so that $D = \gamma E$, where $E$ represents the most recent EPS figure, and noting that we must therefore also have earnings growth of $g=g_D$, it is straightforward to obtain
\[\frac{P}{E}=\frac{(1+g)\gamma}{r-g},\]
as given in \cite{malkiel70}, as the correct $P/E$-ratio for the stock. Although simple, the above relation exhibits a number of features that would be expected given the underlying fundamentals.
First and foremost, as we let $g\nearrow r$, the fair $P/E$-ratio grows without bound. This is in-line with the assertion that investors should be willing to accept higher $P/E$ multiples for higher anticipated earnings growth. In contrast, the larger the value of $r$, corresponding to a higher required rate of return, the lower the fair $P/E$ should be. Higher required rates of return typically indicate a higher level of uncertainty for future cash-flows, as such investors should expect to pay a lower multiple of current earnings.
Finally, as the payout ratio $\gamma$ is increased, so does the $P/E$ value,  due to the greater cash-flows to the investor. However, we should note that the ability to maintain a constant growth rate of $g$ requires a certain level of re-investment, restricting the payout ratio \cite{higgins77}. The Gordon growth model may be generalised to account for a non-constant growth rate over the lifetime of the stock. For example, it is common for immature growth companies to experience rapid earnings growth during the early stages of their development, before settling to a more conservative growth rate as they mature. Such a two stage growth profile can be simply built into the model by suitably adjusting the future cash-flows \cite{SternGGM}. 

Whilst the basic Gordon growth model assumes a constant and equal growth rate for both earnings and dividends, a more general framework is provided by the abnormal earnings growth (AEG) model of Ohlson and Juettner-Nauroth \cite{Ohlson2005}. The model relates the stock price to the following year's expected earnings per share (12 months forward EPS), the short-term growth in EPS (FY-2 versus FY- l), the long-term (asymptotic) growth in EPS and the cost-of-equity capital. Following the derivations of \cite[Corollary 1]{Ohlson2005}, it is possible to obtain an expression for the forward price-earnings ratio which is dependent on the aforementioned factors \cite[Equation 7]{Ohlson2005}. Furthermore, if we impose the assumption of constant and equal growth rates for earnings and dividends, then we recapture the basic Gordon growth model from the AEG.

\section{A growth adjusted $P/E$-ratio}
Having outlined a number of existing value metrics, we now derive our own growth adjusted price-earnings ratio or `GA-$P/E$'. The motivation for the form of the GA-$P/E$ comes from the interpretation of the $P/E$-ratio as a period of time, that being the number of years required for the stocks accrued earnings per share to equal its current price. As such, the GA-$P/E$ bears somewhat of a similarity to the PEG payback period of the previous section. However, through its construction it offers an increased level of precision and a greater ability to differentiate the value offered by stocks.\\

\noindent Let us consider a stock with annual (previous 4 quarters) earnings per share of $E>0$ and an earnings growth rate of $g>-1$, which we assume for now to be known. The total earnings attributable to each stock over the next $N\geq 1$ years are then given by 
\begin{equation}\label{earningstot}
E_N=\sum_{n=1}^{N}E(1+g)^{n}=E(1+g)\frac{(1+g)^{N}-1}{g},
\end{equation}
where the last equality comes as a consequence of the standard result for the first $N$ terms of a geometric series. Relaxing the assumption on $N$ being a positive integer and recalling the interpretation of the standard $P/E$-ratio as the earnings payback period, we equate the above quantity to the stock price $P$ to obtain
\begin{equation}\label{priceeqt}
P=E(1+g)\frac{(1+g)^{N}-1}{g}.
\end{equation}
The $N$ which solves \eqref{priceeqt}, assuming a solution exists, then gives us the required payback period. A simple rearrangement of \eqref{priceeqt} produces
\begin{equation*}
(1+g)^{N}=1+\frac{g}{1+g} \frac{P}{E}.
\end{equation*}
Taking the logarithm of both sides allows us to extract the exponent $N$, and dividing both sides by $\log(1+g)$ then gives
\begin{equation}\label{GA-$P/E$}
N=\frac{\log\left(1+\tfrac{g}{1+g} \tfrac{P}{E}\right)}{\log(1+g)},
\end{equation}
as the number of years required to repay the stock price $P$ through earnings, i.e. the growth adjusted $P/E$-ratio (GA-$P/E$). Having derived our growth adjusted price earnings ratio, we now spend time examining its properties and its dependence on the constituent factors $\tfrac{P}{E}$ and $g$. 

\begin{figure}[h]
\centering
\hspace{0mm}\includegraphics[width=9cm]{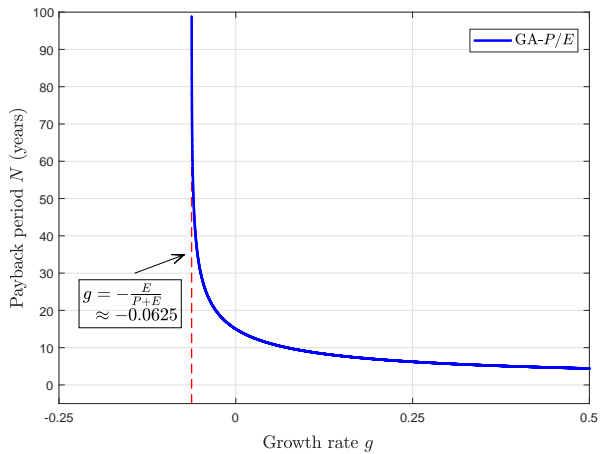}
\caption{GA-$P/E$ Payback period $N$ as growth rate $g$ approaches $-E/(P+E)$.}\label{figure3}
\end{figure}

In our construction of the GA-$P/E$ we assumed the existence of a solution to \eqref{priceeqt}. Implicitly this amounts to an assumption that at some point accrued future earnings will surpass the current stock price $P$. However, it is possible given a sufficiently negative $-1<g<0$, that annual earnings contract sufficiently rapidly that total future earnings amount to less than $P$. Therefore we now derive a lower bound on the growth (contraction) rate $g$ such that we can ensure a solution to \eqref{priceeqt} exists. Let us assume 
that $-1<g<0$, then for a solution to \eqref{priceeqt} to exist we require that
\begin{equation*}
E_\infty=\sum_{n=1}^{\infty}E(1+g)^{n}>P,
\end{equation*}
which, after evaluating the summation, is equivalent to
\begin{equation*}
-E\frac{(1+g)}{g}>P.
\end{equation*}
A simple re-arrangement of the previous inequality provides the following bound on the growth rate $g$:
\begin{equation}\label{bound}
g>-\frac{E}{P+E}.
\end{equation}
If we consider the situation as $g$ decreases towards this bound, then from a financial perspective we would expect it to take an increasingly long period for accrued earnings to repay the initial stock price. Examining the behaviour of \eqref{GA-$P/E$} as we let $g\searrow -E/(P+E)$, we see that $N$ grows without bound as we approach the limit, confirming our intuitive expectations. In Figure \ref{figure3}, we can see this graphically illustrated for the case of $P=15$ and $E=1$. For the given choice of $P$ and $E$, the bound \eqref{bound} computes to approximately -0.0625 and as the growth rate $g$ is reduced towards this level, we observe the GA-$P/E$ $N$ as given in \eqref{GA-$P/E$}, increase sharply.\\

\noindent Another special case we might wish to consider is that of constant earnings, i.e. $g=0$. If we recall, under such an assumption the original price-earnings ratio gives us the payback period for a stock. Therefore, if we substitute $g=0$ in to \eqref{GA-$P/E$} then we would hope to recover $\tfrac{P}{E}$. However, 
if we naively attempt to evaluate \eqref{GA-$P/E$} at $g=0$ then we quickly run in to difficulties as both the numerator and denominator return a value of $0$ for $g=0$. Instead, an application of L’H\^{o}pital’s rule yields the following:
\begin{equation*}
\lim_{g\rightarrow 0}\frac{\log\left(1+\tfrac{g}{1+g} \tfrac{P}{E}\right)}{\log(1+g)}=\lim_{g\rightarrow 0}\frac{(g+1)P}{(g+1)(gE+gP+E)}=\frac{P}{E}.
\end{equation*}

\begin{figure}[h]
\centering
\hspace{5mm}\includegraphics[width=9cm]{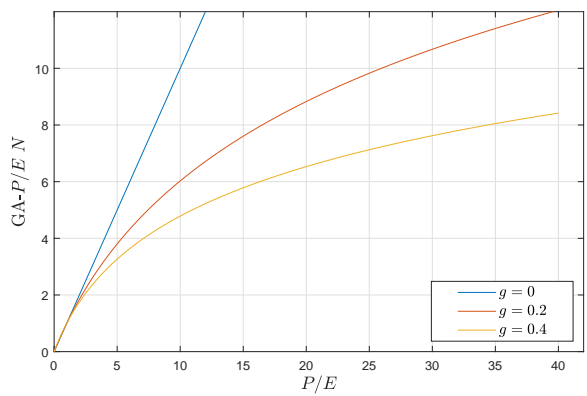}
\caption{GA-$P/E$ payback period $N$ against underlying $P/E$ for various growth rates}\label{figure1}
\end{figure}

\begin{figure}[h]
\centering
\hspace{5mm}\includegraphics[width=9cm]{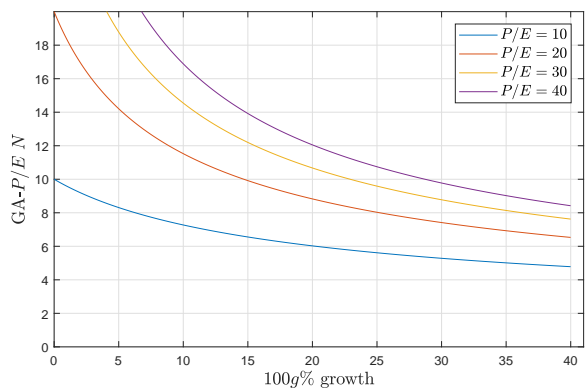}
\caption{GA-$P/E$ payback period $N$ against growth rate $g$ for various underlying $P/E$-ratios}\label{figure2}
\end{figure}

\noindent Hence, in the limit as we let $g\rightarrow 0$, we recover the original price-earnings ratio as we would hope. Having examined some specific cases of $g$, we now consider the behaviour of \eqref{GA-$P/E$} as $g$ is varied more widely. The charts in
Figure~\ref{figure1} show the GA-$P/E$ $N$ plotted against the underlying price-earnings ratio $P/E$ for a selection of growth rates $g$.
\noindent Observing the case $g=0$ (blue line), as discussed above, we see behaviour consistent with the standard $P/E$-ratio, whereby the payback period $N$ equals $P/E$. 
However, as the growth rate $g$ increases, we see that larger initial price-earnings ratios can be supported within a given payback period. In Figure \ref{figure2} we see the GA-$P/E$ plotted against the earnings growth rate for a selection of initial price-earnings ratios. As you would expect, given an initial price-earnings ratio $P/E$, as the growth rate $g$ increases the GA-$P/E$ $N$ decreases.

\section{Empirical study: evaluating the GA-$P/E$}
Having outlined the derivation of our growth adjusted price-earnings ratio and examined some of its properties, we now seek to determine its efficacy as a value metric and predictor of future stock returns. In order to assess this we conducted an empirical study utilising a sorted portfolio methodology. Such studies have found widespread application across empirical finance and in particular have been applied in many studies examining the effectiveness of the traditional price-earnings ratio (or alternatively its inverse, the earnings yield $E/P$), for example \cite{basu77,BASU83,banzgreen86,jaffe89,brooks06}. For an extensive list of such studies the reader may consult the introductory section of \cite{conrad03}. This study proceeds by taking a sample pool of stocks and calculating the GA-$P/E$ for each of the securities within the set. The stocks are then ranked according to their GA-$P/E$ and placed with equal weighting in to quantile portfolios based on this ranking. In this way we end up with a collection of portfolios, ranging from a portfolio containing the lowest GA-$P/E$ stocks to a portfolio consisting of the highest GA-$P/E$ stocks. For each portfolio we compute the monthly returns over the subsequent year. This process of portfolio formation and return tracking is repeated each year over the duration of the study period, although the stocks within the selection pool might vary from year to year as the result of new listings and de-listings. Through this process we obtain a track-record of the historical returns for each of the portfolios. If the GA-$P/E$ is indeed an effective measure of value and predictor of future returns then we would expect to observe an inverse relationship between the ranking of portfolio returns and their GA-$P/E$ ordering, that is lower GA-$P/E$ portfolios providing higher returns and vice versa.

\subsection{Data sources}
The calculation of the GA-$P/E$ and of the subsequent portfolio returns requires the use of both earnings data as well as a record of historical stock prices. For earnings data we utilised the Compustat North American fundamentals dataset, whilst for historical prices we made use of the Center for Research in Security Prices (CRSP) monthly file dataset. The data within the Compustat dataset is provided on a firm by firm basis and is predominantly obtained from SEC filings, whilst the data from CRSP is security specific, with data originating from a number of exchanges. The datasets were then combined using the CRSP/Compustat Merged (CCM) link table. The CCM link table details the linkage histories between firms and securities, allowing us to associate company earnings records with the appropriate securities. For the purposes of our study we restricted our attention to primary listings of common shares on the NYSE, Nasdaq and AMEX over the 25 year period from 1990 to 2015.

Considering specifically the earnings data form Compustat, if we attempt to use reported annual earnings then the fact that companies follow different reporting calendars raises an issue for the annual portfolio construction exercise. For example, if we consider the fiscal year 1996 then we may find this ending as early as June 1996, or as late as May 1997. This misalignment means that whichever date we select to carry out the annual portfolio construction, there will always be a group of firms for which the most recent annual earnings figures are stale. Previous studies, for example \cite{basu77,jaffe89}, get around this issue by considering only those firms whose fiscal year ends in December. However, to avoid further reducing our stock universe, beyond the restrictions made in the upcoming Section \ref{section42}, we instead make use of reported quarterly earnings. Each firm's fiscal year is then redefined to consist of four successive quarters, with the final quarter being the one ending between October and December. In the event that earnings data was unavailable for any of the quarters, then the firm was excluded from the subsequent year's portfolios.

Another issue that we must be aware of is the potential introduction of look-ahead bias in our results. In practice, earnings figures relating to a period are not publicly available immediately following the end of that period. If we were to naively construct our portfolios following the end of our annual fiscal period each December, then we would be utilising data not available to the investor at that time, biasing our results \cite{banzgreen86}. To counteract this we introduced a lag to our study. Since the vast majority of firms report within three months of the earnings period end, we follow \cite{basu77,cookrozeff84,jaffe89} and form our portfolios at the end of March each year, using the month end price and the previous year's earnings to form the GA-$P/E$. 

A further source of potential error in our study is that of survivorship bias, occurring as a result of the way in which firms are added to the Compustat database \cite{Kothari95}. Typically firms are added to Compustat only after they have been operating viably for a number of years, at which point they are introduced with multiple years of historical accounting data, in a process known as `backfilling'. Firms which disappear before reaching such a stage fail to make it onto the database and therefore would be missing from our study. However, there are a number of factors particular to our study which reduce the potential significance of this effect. Firstly, the construction of our GA-$P/E$ measure requires us to attribute an earnings growth rate to each stock under consideration, in our case this is estimated from historical earnings data. The calculation of this historical growth rate is outlined in the following section, however, for now we note that the firm is required to have positive earnings for at least two of its most recent fiscal years. Any firms not satisfying this requirement would automatically be discounted from consideration, hence we might expect a sizeable proportion of the firms missing from Compustat to be ineligible anyway. Furthermore, following 1978, Compustat initiated a major database expansion project, greatly improving their coverage and reducing the significance of the selection biases \cite{Kothari95}. The period of our study occurs after this expansion and therefore the impact of this survivorship bias should be further reduced.

Having covered the addition of new listings to our stock universe, we must also consider how we handle the removal of stocks due to delisting. When a stock within our universe is delisted from its exchange and hence disappears from the CRSP monthly pricing dataset we require a return be attributed to its final month. To compute said return, we ustilised the CRSP monthly stock event delisting dataset, and in particular the delisting return (excluding dividends) data item. Although such delistings may occur at any point throughout the month, the realisation of any return is attributed to the end of the month in which it occurs. In the case of a stock disappearing from the pricing dataset without a delisting return being available, we assume a complete loss for the investor and attribute a return of -1.

\subsection{Estimating growth}\label{section42}
The calculation of the GA-$P/E$ measure \eqref{GA-$P/E$} requires us to attribute an earnings growth rate $g$ to each stock in any given year. Within our study, these rates were approximated from historical earnings data in the following manner. Let us suppose we are calculating the GA-$P/E$ for a given stock as part of the portfolio formation process during the calendar year $n$. As such, the most recent earnings figure will be derived from the four quarter period ending in the final quarter of the calendar year $n-1$, which we denote by $E_{n-1}$. Now assuming we wish to estimate the growth over a historical period of $k$ years, 
i.e. the growth between $E_{n-k-1}$ and $E_{n-1}$, then we estimate the growth rate $g$ for the year $n$ as follows:

\begin{equation}\label{growth_rate}
g=\left(1+\frac{E_{n-1}-E_{n-k-1}}{E_{n-k-1}}\right)^{\tfrac{1}{k}}-1.
\end{equation}
\newline
\noindent In our empirical study we utilised a range of period lengths $k(=1,2,3)$ when computing our growth estimates. We also note that the requirement for historical earnings from which to estimate the growth rate necessitates a longer history for our earnings data than for our price data. Given that the period of our study commences with the first portfolio formation in March 1990, and that we use up to a 3 year window to estimate the growth rate $g$, we therefore require 4Q earnings figures dating as far back as October 1986.
\begin{figure}[h]
\centering
\hspace{5mm}\includegraphics[width=9cm]{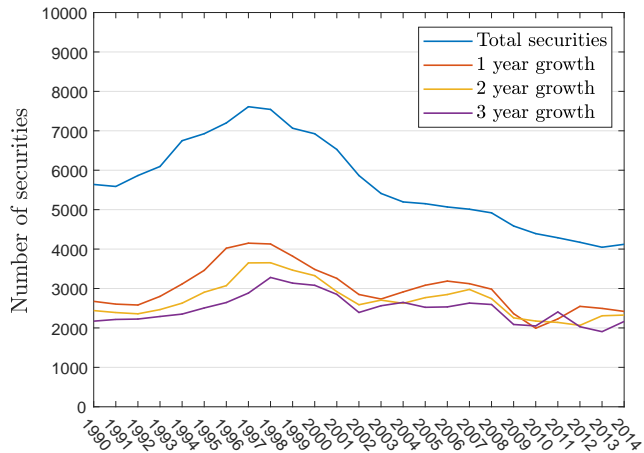}
\caption{Total number of matched securities and the number meeting the earnings conditions for $k=1,2,3$ over the period 1990-2014}
\label{earncond}
\end{figure}

\noindent The calculation \eqref{growth_rate} requires that both the earnings figures $E_{n-k-1}$ and $E_{n-1}$ are positive. Therefore, when forming our portfolios in year $n$ we must exclude any stock which does not meet this requirement. The charts in Figure \ref{earncond} depict the scale of this reduction throughout the period of the study. The uppermost blue line plots the number of securities resulting from the matching procedure and represents the maximum extent of our stock universe. The remaining lines show the number of securities left after we have imposed the restriction that both $E_{n-k-1}$ and $E_{n-1}$ be positive, where we have considered the values $k=1,2,3$.

We would expect company earnings to display a degree of autocorrelation, with firms making a profit (loss) one year being more likely to make a profit (loss) in subsequent years. The effect of this serial correlation is likely to be strongest in consecutive years and diminish as the years in question become further apart. The impact of this may be observed in Figure \eqref{earncond}, where the number of securities meeting the earnings condition is generally highest for $k=1$ and the lowest for $k=3$. 

\subsection{Handling cases with an infinite GA-$P/E$}

\noindent As determined earlier, when the growth rate $g$ approaches and drops below the value \eqref{bound} the GA-$P/E$ \eqref{GA-$P/E$} grows without bound, as the totality of projected future earnings falls below the current stock price. This raises an issue regarding how to handle these `infinite GA-$P/E$' stocks, should they be excluded from the study, or else if we choose to include them, which portfolio should they be allocated to? The following figure (Figure \eqref{proportions}) depicts the proportion of those stocks satisfying the positive earnings requirements, which produce a finite and GA-$P/E$s. Over the duration of the study period, we observe that a non-negligible proportion of stocks (generally around 25-30\% per year) have an infinite GA-$P/E$. Therefore, if we were to remove them from the study then we would be significantly reducing our sample size, beyond the cuts made on the basis of positive earnings. Preferring to avoid the further reduction of our stock universe, we proceed to outline a means by which to rank the infinite GA-$P/E$ stocks within our framework. 
\begin{figure}[H]
\centering
\hspace{5mm}\includegraphics[ width=9cm]{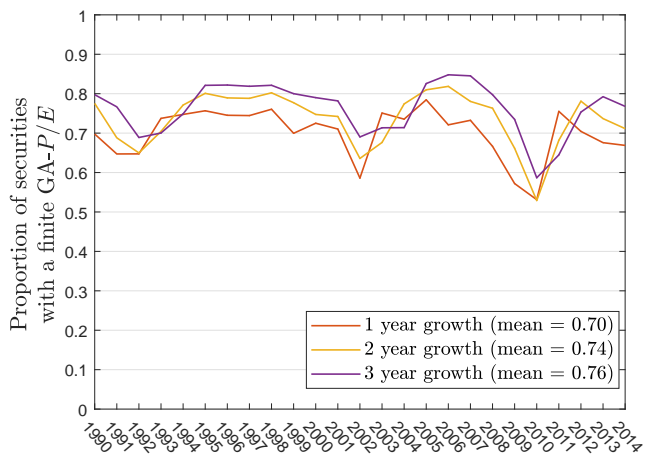}
\caption{Proportions of stocks with finite and infinite GA-$P/E$s when growth is calculated over 1,2 and 3 years}
\label{proportions}
\end{figure}

\noindent  Clearly we should consider stocks with an infinite GA-$P/E$ to lie at the costly end of our value spectrum, and as such they should be allocated to the portfolio consisting of the highest GA-$P/E$ stocks. However, it is not necessarily the case that the quantile portfolios will always be sufficiently large to allow the inclusion of all infinite GA-$P/E$ stocks within a single portfolio. In fact, with typically around 25\% of stocks having an infinite GA-$P/E$, there is likely to be some overspill from the highest GA-$P/E$ portfolio for at least some of the years. One possible solution might see us allocate all infinite GA-$P/E$ stocks to a standalone highest GA-$P/E$ portfolio, before separating the remaining finite GA-$P/E$ stocks into a number of other portfolios. A similar approach was adopted by \cite{banzgreen86}, whilst studying the $E/P$ ratio. The authors placed all stocks with negative earnings, hence a negative $E/P$, in a single portfolio before allocating the remaining stocks into equally sized portfolios based on their $E/P$. However, in our case, given that the proportion of stocks with an infinite GA-$P/E$ is not constant, and can be seen from Figure \eqref{proportions} to exceed 40\% in some years, such an approach would lead to significant discrepancies in portfolio size for some of years. Therefore, we require a means of ranking infinite GA-$P/E$ stocks amongst themselves in order to determine which of them to attribute to the highest GA-$P/E$ portfolio, and which if any to allocate to lower GA-$P/E$ portfolios. In order to do so, we consider the proportion of the stock price $P$ that is projected to be paid back. In the case that \eqref{bound} is not satisfied, this payback proportion is given by

\begin{equation}
\label{ppprop}
\frac{E_\infty}{P}=\frac{\sum_{n=1}^{\infty}E(1+g)^{n}}{P}=\frac{E(1+g)}{-gP}.
\end{equation}

\noindent In contrast to the GA-$P/E$ measure where a lower figure signifies greater value, the higher \eqref{ppprop} is calculated to be, the greater the value offered by the the underlying stock. In order to remain consistent with the GA-$P/E$ we therefore take the reciprocal of the above ratio to get $P/E_{\infty}$. Furthermore, if $N_{max}$ signifies the highest finite GA-$P/E$ observed in a given year, then for all the infinite GA-$P/E$ stocks that year, we add $N_{max}$ to the inverted ratio $P/E_{\infty}$ to give the following measure:
\[N^{*}=N_{max}+P/E_{\infty}= N_{max}-\frac{gP}{E(1+g)}.\]
\noindent By using the standard GA-$P/E$ for those stocks which return a finite value of \eqref{GA-$P/E$}, and $N^{*}$ for those stocks registering an infinite GA-$P/E$, we obtain a value measure which facilitates a coherent ranking across both finite and infinite GA-$P/E$ stocks. However, we note that the use of $N^{*}$ is only to allow a ranking to be taken, the measure $N^{*}$ has no intrinsic significance and the underlying stocks all possess an infinite GA-$P/E$.

\subsection{Empirical results}\label{empirical}
Following the approach outlined, we conducted our empirical study over the period from 1990 to 2015, with portfolio construction occurring at the end of each March from 1990 to 2014. During each construction event we formed five quintile portfolios, labeled $P_1$ to $P_5$, corresponding with a low to high GA-$P/E$ ordering. The choice of five portfolios is in itself arbitrary, however it matches the number selected by Basu \cite{basu77}. From Figure \ref{earncond}, we can see that the number of stocks within our pool typically varies between 2000 and 4000 depending upon the year and the period over which growth is calculated. Therefore, upon construction we would expect our portfolios to contain between 400 and 800 individual stocks. Furthermore, if we consider the proportion of stocks displaying an infinite GA-$P/E$  (Figure \ref{proportions}), then it allows us to apportion the majority of such stocks to the highest GA-$P/E$ portfolio, whilst retaining a reasonable number of portfolios and hence a range of typical GA-$P/E$ values. This process was followed using a one, two and three year window for estimating our growth rates, with the corresponding restrictions of our stock universe as per Figure \ref{earncond}. The following table (Table~\ref{stats1}) provides a summary of the typical value of some key variables for the resulting GA-$P/E$ sorted portfolios. The values given in the table are obtained by averaging the median values (for individual years) over the duration of the study.

\begin{table}[H]
\begin{center}
\begin{threeparttable}
\begin{tabular}{ccccccc}\toprule
&Portfolio & $P_1$ & $P_2$ & $P_3$ & $P_4$ & $P_5$   \\
\toprule

\rule{-3pt}{10pt}
1 year growth&GA-$P/E$  & 2.46&	5.13  & 9.05 & 18.66\tnote{*} &	NA  \\
window&$P/E$ & 12.61 &	15.28 &	16.61 &	16.73 & 29.57 \\
&Growth rate (\%) & 144.32 & 38.98 &	14.65 &	-5.84 &	-48.92 \\
\midrule
\rule{-3pt}{10pt}
2 year growth & GA-$P/E$   & 3.17 &	5.84 &	9.46&	18.52\tnote{**}	& NA  \\
window&$P/E$  & 11.68 &	14.91 &	16.65 &	17.82 &	31.31 \\
&Growth rate (\%) & 80.35 &	29.04 &	13.01 &	-2.02&	-32.17 \\
\midrule
\rule{-3pt}{10pt}
3 year growth &GA-$P/E$  & 3.67 & 6.32 & 9.43 &	19.23\tnote{***} &	NA  \\
window&$P/E$ & 11.23 &	14.76 &	16.67 &	18.59	&31.65 \\
&Growth rate (\%) & 57.93 &	24.16 &	11.82 &	0.14 &	-23.76\\
\bottomrule
\end{tabular}
\begin{tablenotes}
\footnotesize \item[*] Excluding ten years in which the median GA-$P/E$ was infinite. ${}^{**}$ Excluding seven years in which the median GA-$P/E$ was infinite. ${}^{***}$ Excluding four years in which the median GA-$P/E$ was infinite. NA signifies that the median GA-$P/E$ was infinite in all years.
\end{tablenotes}
\end{threeparttable}
\end{center}
\caption{Summary of statistics for the GA-$P/E$ portfolios when growth was estimated using a one, two and three year window.}
\label{stats1}
\end{table}

\noindent Examining the statistics given in Table~\ref{stats1}, we can see that in each case, the lowest GA-$P/E$ stocks tend to be those with the lowest $P/E$-ratios and the highest earnings growth, with the opposite being true for the highest GA-$P/E$ stocks. Although this might be expected from the form of the GA-$P/E$, it does somewhat run contrary to the distinction between value and growth stocks and the idea that the cost of high earnings growth is inevitably a high $P/E$-ratio. Considering the combination of extremely high/low growth rates coupled with low/high $P/E$-ratios, it is likely that many of the extreme GA-$P/E$ stocks are those that have experienced a significant change in their earnings over the short-term which has not been matched by a comparable change in share price. Comparing the statistics across the different growth windows, it is noticeable that the growth rates associated with our end portfolios $P_1$ and $P_5$ are significantly moderated as the window length increases. This might have been expected, since the likelihood of a firm recording significant changes in earnings over a second or third consecutive year diminishes, as compared to the likelihood of observing a single year of high earnings growth/contraction.

Having formed the portfolios $P_1$ to $P_5$ using one, two and three year growth windows, we average the holdings across the three realisations of each portfolio, to create a single set of five portfolios. This is done in order to mitigate the noise within the estimated earnings growth rates, which will often display high levels of variability from year to year \cite{Little62}. By averaging across the three instances, we place a greater weight upon those stocks which have consistently displayed a growth rate $g$, which when combined with the associated $P$ and $E$, produce a GA-$P/E$ that would place the stock in the given portfolio.

As explained in Section \ref{section42}, the requirement for positive earnings for the end years of our growth windows restricts the pool of stocks from which we may select the portfolios. Given the pools associated with one, two and three year windows, we also form three sets of quintile portfolios based upon a ranking of $P/E$-ratios. As with the GA-$P/E$ sorted portfolios, we average the holdings across the three instances of each portfolio, in order to obtain one set of five $P/E$ sorted portfolios.

The average annual returns of our five GA-$P/E$ portfolios can be found in the first row of Table~\ref{returns1}. Listed alongside these returns, we have included the those of the analogous $P/E$ sorted portfolios. 

\begin{table}[H]
\begin{center}
\begin{threeparttable}
\begin{tabular}{cccccc} \toprule
\rule{0pt}{12pt} &  \multicolumn{5}{c}{Average annual rate of return\footnotemark }  \\ \hline
\rule{-2pt}{12pt} Portfolio & $P_1$ & $P_2$ & $P_3$ & $P_4$ & $P_5$ \\ \midrule
\rule{-3pt}{10pt}
GA-$P/E$ sort       & 0.1982	& 0.1660 & 0.1367 & 0.1336 & 0.1580\\
$P/E$ sort  & 0.1889 & 0.1569 & 0.1377 & 0.1424 & 0.1598 \\
\midrule
\rule{-2pt}{10pt}Difference & \cellcolor{green!45}0.0093 & \cellcolor{green!45}0.0091 & \cellcolor{red!45}-0.0010 &\cellcolor{red!45} -0.0087 &\cellcolor{red!45} -0.0017 \\
\bottomrule
\end{tabular}
\begin{tablenotes}
\vspace{2mm}
\item \footnotesize\crule[green!45]{.25cm}{.25cm}\hspace{1mm}signifies that the return from the GA-$P/E$ portfolio exceeds that of the corresponding $P/E$ portfolio, whilst\hspace{1mm} \crule[red!45]{.25cm}{.25cm}\hspace{1mm}signifies the return from the $P/E$ portfolio exceeds that of the corresponding GA-$P/E$ portfolio.  
\end{tablenotes}
\end{threeparttable}
\end{center}
\caption{Average annual returns for the averaged GA-$P/E$ and $P/E$ portfolios.}
\label{returns1}
\end{table}
\footnotetext{The period of the study covers 300 months. If $r_1,\ldots, r_{300}$ denote the associated monthly returns, then the average annual rate of return is given by $\prod_{j=1}^{300}\left(1+r_j\right)^{12/300} -1$. }

\begin{figure}[H]
\centering
\hspace{5mm}\includegraphics[width=9cm]{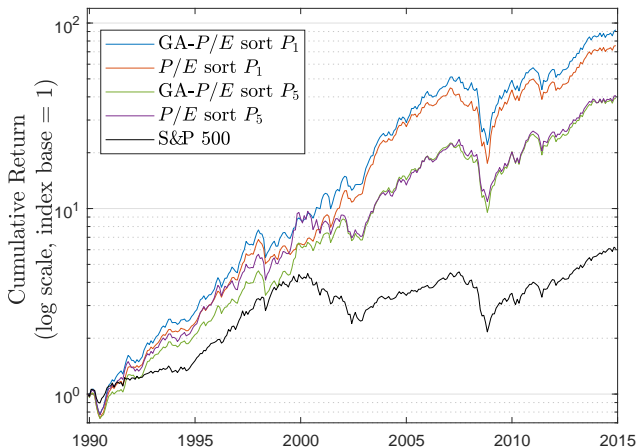}
\caption{Cumulative performance over the course of the study (1990-2015) of portfolios $P_1$ and $P_5$ based upon GA-$P/E$ and $P/E$ sorts, compared with the S\&P 500.}
\label{performance}
\end{figure}

\noindent From the briefest examination of Table \ref{returns1} it is clear that in general the lower GA-$P/E$ portfolios provide the highest average returns, with $P_1$ and $P_2$ providing average annual returns of 19.82\% and 16.60\% respectively. The only outlier to the trend being $P_5$, whose return of 15.80\% exceeds that of both $P_3$ and $P_4$. However, it is not unexpected that the ordering of returns is not perfectly monotonic, with similar deviations having been observed in many studies of the $P/E$-ratio. For example in \cite{basu77}, the highest $P/E$ (lowest $E/P$) portfolios were found to outperform a number of the middle ranking portfolios, whilst both \cite{jaffe89, fama92} comment on a U-shaped returns profile. Indeed, examining the returns from our $P/E$ ranked portfolios we observe a more pronounced U-shaped profile, with the highest $P/E$ portfolio $P_5$ delivering the second highest return, whilst the lowest returns are generated by the middle ranked portfolio $P_3$. 
However, once again the highest returns are provided by the value portfolio $P_1$. Comparing the level of returns between the GA-$P/E$ and $P/E$ portfolios, for the value portfolios ($P_1$ and $P_2$), the GA-$P/E$ noticeably outperforms the $P/E$, providing an additional 0.9\% annually in both cases. Whereas at the other end, for the portfolios $P_4$ and $P_5$, the returns for the GA-$P/E$ can be seen to be below those for the $P/E$. The cumulative effect of the return differential between the low and high GA-$P/E$ portfolios $P_1$ and $P_5$ can be seen in Figure \ref{performance}. The chart also details the cumulative performance of the low and high $P/E$ portfolios, allowing us not only to observe the difference between these two portfolios, but also to compare between analogous GA-$P/E$ and $P/E$ portfolios.

At first glance, the returns in Table \ref{returns1} and Figure \ref{performance} may appear rather high especially in comparison to the S\&P 500 which gave an average annual return of 7.49\% over the same period. Considering this further, there are a number of reasons which might explain the strong performance. The first point to note is that in order to estimate the earnings growth rate, we were required to drop those firms which recorded a loss in either the year starting or ending our growth window. The firms that are left are those with a more robust profitability record. Secondly, by including all primary listings on the NYSE, Nasdaq and AMEX we would expect the average size of a firm within our stock universe to be somewhat smaller than the typical S\&P 500 firm. Furthermore, before averaging holdings over the different growth windows, each firm within our base portfolios initially receives an equal weighting. As such, following the averaging procedure, the maximum difference in weighting that we might observe between two stocks within the same portfolio is approximately 3 to 1. In contrast, the S\&P 500 employs a weighting system based upon market capitalisation, and therefore we observe significantly larger differences in weightings (greater than 3 to 1), in favour of larger firms over smaller firms. As a result, the significance of smaller firms within our portfolios is increased in comparison to an equivalent market capitalisation weighted portfolio, or indeed the S\&P 500. Since both high profitability and smaller firm size are factors which correlate with higher stock returns \cite{fama15}, it is unsurprising that our portfolios outperform the S\&P 500 by a margin.

Having observed that the returns of the lower GA-$P/E$ portfolios were greater than those of the high GA-$P/E$ portfolios over our test period, we now investigate whether these differences in returns are statistically significant. We also tested the differences in returns between the $P/E$ portfolios so as to allow comparison. We conducted paired sample, two-tailed $t$-tests between the monthly returns $\left \lbrace R_i \right\rbrace$ and $\left \lbrace R_j \right \rbrace$ of portfolios $P_i$ and $P_j$, for both the GA-$P/E$ and the $P/E$ portfolios. The resulting $p$-values may be found in Table~\ref{pvalues1}.

\begin{table}[H]
\begin{center}
\begin{threeparttable}
\begin{tabular}{|c|cccc||c|cccc|}
\hline
 \multicolumn{5}{|c||}{\rule{-2pt}{12pt}  $p$-values for GA-$P/E$ monthly returns }    &     \multicolumn{5}{c|}{\rule{-2pt}{12pt}  $p$-values for $P/E$ monthly returns }   \\ \hline
\rule{-2pt}{12pt} & $P_2$     & $P_3$     & $P_4$       & $P_5$     &    & $P_2$     & $P_3$     & $P_4$     & $P_5$     \\ \hline
\rule{-3pt}{12pt} \hspace{-2mm}$P_1$\hspace{-2mm} & 6$\times 10^{-4}$ \tnote{***}\hspace{0.5mm} \hspace{0.5mm}  &
2$\times 10^{-5}$ \tnote{***}\hspace{0.5mm}\hspace{0.5mm} & 2$\times 10^{-6}$\hspace{0.5mm}\tnote{***}\hspace{0.5mm}\hspace{0.5mm} & \hspace{0.5mm}0.0012\hspace{0.5mm}\tnote{**}\hspace{0.5mm}   & \hspace{-2mm}$P_1$\hspace{-2mm} & 0.0030\hspace{0.5mm}\tnote{**}\hspace{1mm}  & 2$\times 10^{-4}$\hspace{0.5mm}\tnote{***}\hspace{0.5mm}\hspace{0.5mm} & 0.0044\hspace{0.5mm}\tnote{**}\hspace{1mm}  & \hspace{0.5mm}0.2830\hspace{.1mm}\\
\rule{-3pt}{10pt} $P_2$ & \cellcolor{gray!25}        & 5$\times 10^{-5}$\hspace{0.5mm}\tnote{***}\hspace{0.5mm}\hspace{0.5mm}& 1$\times 10^{-4}$\hspace{0.5mm}\tnote{***}\hspace{0.5mm}              & \hspace{-1mm}0.6632         & $P_2$ & \cellcolor{gray!25}        & 0.0117\hspace{0.5mm}\tnote{*}\hspace{1mm}                       &\hspace{-1mm} 0.2964 &\cellcolor{yellow!45} 0.6054 \\
\rule{-3pt}{10pt} $P_3$& \cellcolor{gray!25}        & \cellcolor{gray!25}        & 0.6789             & \cellcolor{yellow!45}\hspace{-1mm}0.0681           & $P_3$ &  \cellcolor{gray!25}       & \cellcolor{gray!25}        &\cellcolor{yellow!45}\hspace{-1mm} 0.4202 &\cellcolor{yellow!45} 0.1184 \\
\rule{-3pt}{10pt} $P_4$ & \cellcolor{gray!25}        & \cellcolor{gray!25}        & \cellcolor{gray!25}   & \cellcolor{yellow!45}\hspace{0.65mm}0.0104 \hspace{-2mm} \tnote{*}\hspace{1mm}           & $P_4$ & \cellcolor{gray!25}        &  \cellcolor{gray!25}       & \cellcolor{gray!25}        &\cellcolor{yellow!45} 0.0967 \\ \hline
\end{tabular}
\begin{tablenotes}
\item ${}^{*,}$  ${}^{**,}$  ${}^{***}$  \footnotesize denote significance at the 5\%, 1\% and 0.1\% levels respectively.
\item \footnotesize\crule[yellow!45]{.25cm}{.25cm}\hspace{1mm} signifies that the observed return differential was negative, counter to the proposed monotonic ordering.
\end{tablenotes}
\end{threeparttable}
\end{center}
\caption{$p$-values for the differential $\bar{R}_i-\bar{R}_j$ in average monthly returns between portfolio $P_i$ ($i^{th}$ row) and portfolio $P_j$ ($j-1^{th}$ column) using a two-tailed $t$-test. Portfolios formed on the basis of GA-$P/E$ (left) and $P/E$-ratio (right).}
\label{pvalues1}
\end{table}

\noindent Examining the results of Table \ref{pvalues1}, we see a greater number of statistically significant results when comparing the monthly returns between the GA-$P/E$ sorted portfolios than we do for the $P/E$ sorted portfolios. Most notably, when we test the returns of the lowest GA-$P/E$ portfolio $P_1$ against those of the other portfolios, we obtain a significant result in each case. However, this is not the case with the $P/E$ sorted portfolio, where the difference in returns between $P_1$ and $P_5$ is not significant. Furthermore, in those cases where the returns differential for both the GA-$P/E$  and the $P/E$ portfolios are significant, we invariably find that the result for the GA-$P/E$ shows a higher level of significance. This evidence further supports the hypothesis that the GA-$P/E$ is predictive of future stock returns, and when selection is restricted to stocks with positive earnings, the predictive strength of the GA-$P/E$ exceeds that of the $P/E$-ratio.

We now examine the returns of the portfolios in more depth, adjusting the returns to account for their observed volatility, and also determine the portfolios' exposure to some common risk factors. In order to assess the risk-adjusted returns of our portfolios we utilise the Sharpe ratio. Developed by William Sharpe in \cite{sharpe66} and later revised in \cite{Sharpe94}, the Sharpe ratio measures the expected excess return earned by an asset per unit of volatility, where the excess return is measured against the risk-free rate. In our analysis, when computing the Sharpe ratio we utilise monthly returns \footnotemark data and take the 3-month U.S. Treasury as a proxy for the risk-free asset. We also measure the exposure of the portfolios' returns to the standard risk factors identified in Fama and French's \cite{fama93} three factor model (market risk factor, HML value risk factor and SMB size risk factor). In our assessment of these exposures we utilise the monthly returns of the S\&P 500 as a proxy for the market returns and once again take the 3-month U.S. Treasury as a proxy for the risk-free asset. The data for both the monthly S\&P and 3-month U.S. Treasury returns were obtained from WRDS, as were the monthly values of the HML and SMB Fama-French factors.

\footnotetext{Whilst the computed Sharpe ratios are given on a monthly basis, we can approximate their annual equivalents by multiplying the resulting figure by $\sqrt[]{12}$, for further details see \cite{lo02}. For reference, following the same calculation, the S\&P 500 gives a Sharpe ratio of $0.1068$ over the period of our study.}

\begin{table}[H]
\begin{center}
\begin{threeparttable}
\begin{tabular}{ccccccc} \cline{1-6}  
\multicolumn{6}{c}{Performance measures and factor coefficients for GA-$P/E$ sorted portfolios }  \\ \hline
\rule{-2pt}{12pt}  Portfolio& $P_1$ & $P_2$ & $P_3$ & $P_4$ & $P_5$   \\ \hline
\rule{-3pt}{10pt}
Sharpe ratio\footnotemark[2]       & 0.2677 & 0.2535 & 0.2297 & 0.2191 & 0.2253	 \\
 \hline
 \vspace{-3mm}  \\

alpha $\alpha$   & 0.0076	& 0.0058 & 0.0040 & 0.0035 & 0.0047	 \\
  & (6.8917)	& (6.8820) & (5.8947) & (6.4153) & (6.0148)	\vspace{1mm}   \\

market beta $\beta$  & 0.9741 & 0.8674 & 0.8040 & 0.8140 & 0.8990  \\
  & (37.3335)	& (43.6491) & (50.3025) & (62.6672) & (48.3175)\vspace{1mm}  	 \\

 
value HML $b_{HML}$   & 0.3668	& 0.3088 & 0.2946 & 0.3334 & 0.4020\\
 & (9.6976)	& (10.7178) & (12.7132) & (17.7054) & (14.9026)\vspace{1mm}  	 \\

size SMB $b_{SMB}$   & 0.8400	& 0.6882  & 0.5168 & 0.5997 & 0.8688	 \\
 & (24.1520)\hspace{2mm}	& (25.9816)\hspace{2mm} & (24.2577)\hspace{2mm} & (34.6341)\hspace{2mm} & (35.0273)\vspace{1mm}  	 \\
$R^2$  & 0.8747 & 0.9010 & 0.9162 & 0.9474 & 0.9264 \vspace{1mm}   \\

 \hline
 \end{tabular}
 \end{threeparttable}
\end{center}
\caption{Performance measures and factor coefficients under Fama-French 3 factor model for GA-$P/E$ sorted portfolios.  The values in parentheses below the regression coefficients are the relevant $t$-statistics.}
\label{perfGAPE}
\end{table}

\begin{table}[H]
\begin{center}
\begin{threeparttable}
\begin{tabular}{ccccccc} \cline{1-6} 
\multicolumn{6}{c}{Performance measures and factor coefficients for $P/E$ sorted portfolios }  \\ \hline
\rule{-2pt}{12pt}  Portfolio& $P_1$ & $P_2$ & $P_3$ & $P_4$ & $P_5$   \\ \hline
\rule{-3pt}{10pt}
Sharpe ratio\footnotemark[2]       & 0.2659 & 0.2567 & 0.2248 & 0.2185 & 0.2116	 \\
 \hline
 \vspace{-3mm}  \\

alpha $\alpha$   & 0.0066	& 0.0050 & 0.0038 & 0.0043 & 0.0057	 \\
  & (5.8681)	& (6.5333) & (5.2290) & (5.7926) & (6.8796)	\vspace{1mm}   \\

market beta $\beta$  & 0.9063 & 0.7922 & 0.8071 & 0.8788 & 0.9854  \\
  & (34.1122)	& (43.8365) & (47.1265) & (50.0248) & (50.5249)\vspace{1mm}  	 \\

 
value HML $b_{HML}$   & 0.6346	& 0.4868 & 0.3822 & 0.2009 & -0.0193\\
 & (16.4766)	& (18.5807) & (15.3924) & (7.8895) & (-0.6818)	\vspace{1mm}   \\

size SMB $b_{SMB}$   & 0.7768	& 0.6223  & 0.5902 & 0.6332 & 0.8754	 \\
 & (21.935)\hspace{2mm}	& (25.9816)\hspace{2mm} & (24.2577)\hspace{2mm} & (34.6341)\hspace{2mm} & (35.0273)\vspace{1mm}  	 \\
$R^2$  & 0.8559 & 0.9024 & 0.9105 & 0.9200 & 0.9334 \vspace{1mm}   \\

 \hline
 \end{tabular}
 \end{threeparttable}
\end{center}
\caption{Performance measures and factor coefficients under Fama-French 3 factor model for $P/E$ sorted portfolios.  The values in parentheses below the regression coefficients are the relevant $t$-statistics.}
\label{perfPE}
\end{table}

\noindent Examining Table \ref{perfGAPE} and the Sharpe ratios for the GA-$P/E$ sorted portfolios, we see a similar U-shaped profile across the portfolios (as observed in Table \ref{returns1}) once the returns are adjusted for risk. The value portfolio $P_1$ gives the highest risk adjusted return, with the Sharpe ratio decreasing as we move down the value range through to portfolio $P_4$. However, once again $P_5$ proves an exception to the monotonic ordering, although after adjusting for risk, the returns from $P_5$ now rank $4^{th}$ as opposed to $3^{rd}$ previously. In comparison, if we examine the risk adjusted returns of the $P/E$ sorted portfolios  from Table \ref{perfPE}, then we can see  a strict monotonic ranking as we go from $P_1$ through to $P_5$, matching the results of \cite{basu77}. Nonetheless, the risk adjusted return for the GA-$P/E$ sorted $P_1$ exceeds that of the $P/E$ sorted $P_1$.

Examining the  GA-$P/E$ sorted portfolios' exposure to the risk factors (Table \ref{perfGAPE}), we see that at least some of  $P_{1}$'s additional return may be explained by higher exposures to the market risk factor ($\beta=0.9741$, $1^{st}$ overall), value risk factor ($b_{HML}=0.3668$, $2^{nd}$ overall) and the size risk factor ($b_{SMB}=0.8400$, $2^{nd}$ overall). However, these exposures do not explain all of the additional returns as evidenced by $P_1$ having the highest (monthly) alpha value ($\alpha=0.0076$), which measures the average abnormal return not explained by the three standard risk factors. Contrasting the exposures for the two $P_{1}$'s, we see that the  GA-$P/E$ derived $P_1$ has a higher exposure to the general market risk factor (higher $\beta$) and also the size risk factor (higher $b_{SMB}$), than does the $P/E$ sorted $P_1$. On the other hand, as we might expect, the portfolio $P_1$ originating from the $P/E$ sort displays a significantly higher exposure to the value risk factor (higher $b_{HML}$), than the GA-$P/E$ sorted $P_1$. Overall, exposure to these standard risk factors is insufficient to explain the additional returns from the GA-$P/E$ portfolio $P_1$, as evidenced by its higher alpha intercept. 

More widely it is interesting to compare the pattern of $b_{HML}$ coefficients from the two sets of portfolios. Looking at the $P/E$ sorted portfolios (Table \ref{perfPE}), we see a very clear decrease in the exposure to the value factor as we go from $P_1$ to $P_5$, with $P_5$ having a negative $b_{HML}$ coefficient. The $P/E$-ratio is a traditional value metric much like the $B/M$-ratio which is used to construct the value factor risk premium, as such it is unsurprising that the portfolios exhibit increasing $b_{HML}$ exposure as we go from the high $P/E$ non-value portfolio $P_5$ through to the low $P/E$ value portfolio $P_1$. In contrast, the GA-$P/E$ portfolios produce a more uniform pattern of $b_{HML}$ coefficients, indicating a more even exposure to the traditional value risk factor across the portfolios. This suggests, that by adjusting our value measure for growth, the GA-$P/E$ finds `true value' across the traditional value spectrum.

We noted that at least some of the additional returns for the GA-$P/E$ portfolio $P_1$ might be attributed to its higher exposure to the size risk factor. The size-effect was first noted by Banz \cite{banz81}, and has now been accepted as a standard risk factor in the asset pricing literature \cite{fama93,fama15,carhart97}. Whilst it would seem plausible that smaller firms experience greater risk in general than their larger counterparts, specific explanations include the lower liquidity of smaller stocks \cite{Amihud86, liu06}, and also increased uncertainty owing to a reduced availability of information about small firms \cite{zhang06}. In any event, evidence suggests that these effects are largely restricted to the smallest of publicly traded firms \cite{Horowitz2000, fama08}, and that by removing these microcap stocks we can mitigate the effect. Therefore, in order to test the GA-$P/E$ further, we restrict our stock universe to the largest $x\%$ of firms by market capitalisation on NYSE, Nasdaq and AMEX, prior to carrying out the same procedure as before. We then reduce the percentage $x\%$ of firms retained, and record the average annual return, the Sharpe ratio and $\alpha$ of the $P_1$ portfolios resulting from both a GA-$P/E$ and a $P/E$ sort. The charts below (Figure \ref{alpha slope}) detail the results of this process.

\begin{figure}[H]
\begin{minipage}[b]{0.45\linewidth}
\centering
\hspace{5mm}\includegraphics[width=8.1cm]{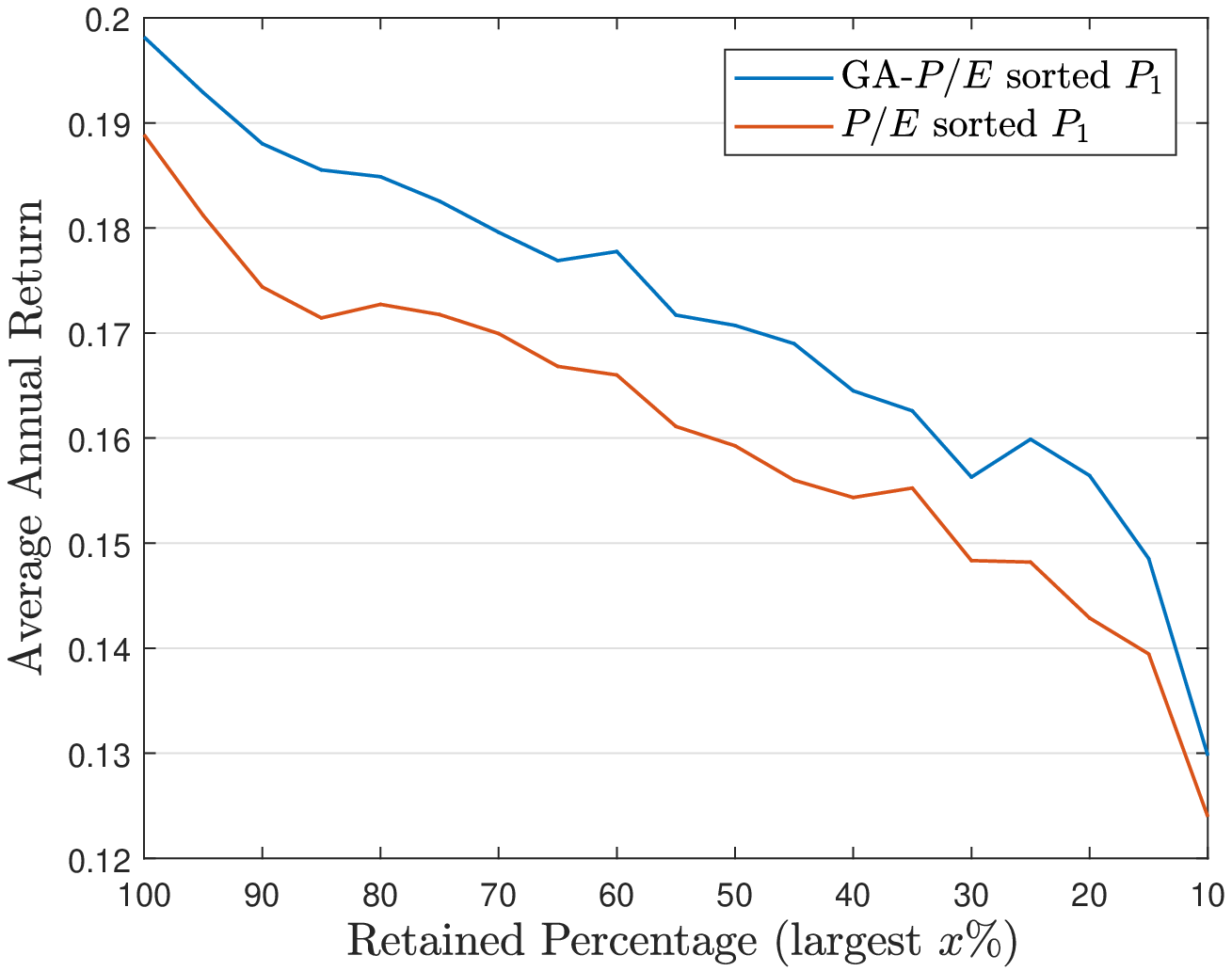}\\\hspace{0mm}\\
\end{minipage}
\hspace{0.5cm}
\begin{minipage}[b]{0.45\linewidth}
\centering
\hspace{5mm}\includegraphics[width=8cm]{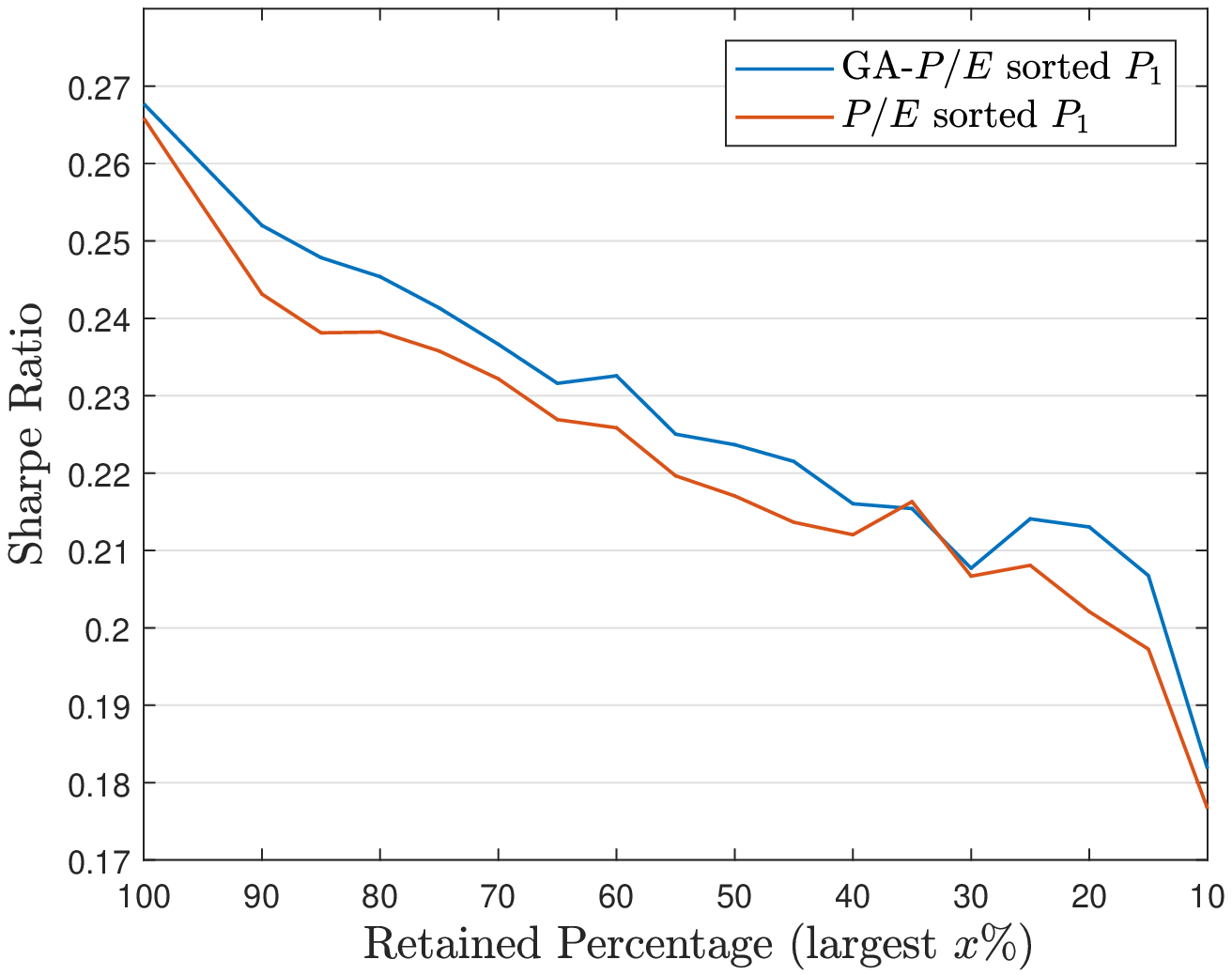}\\\hspace{0mm}\\
\end{minipage}
\end{figure}
\vspace{-1cm}
\begin{figure}[H]
\centering
\hspace{5mm}\includegraphics[width=8cm]{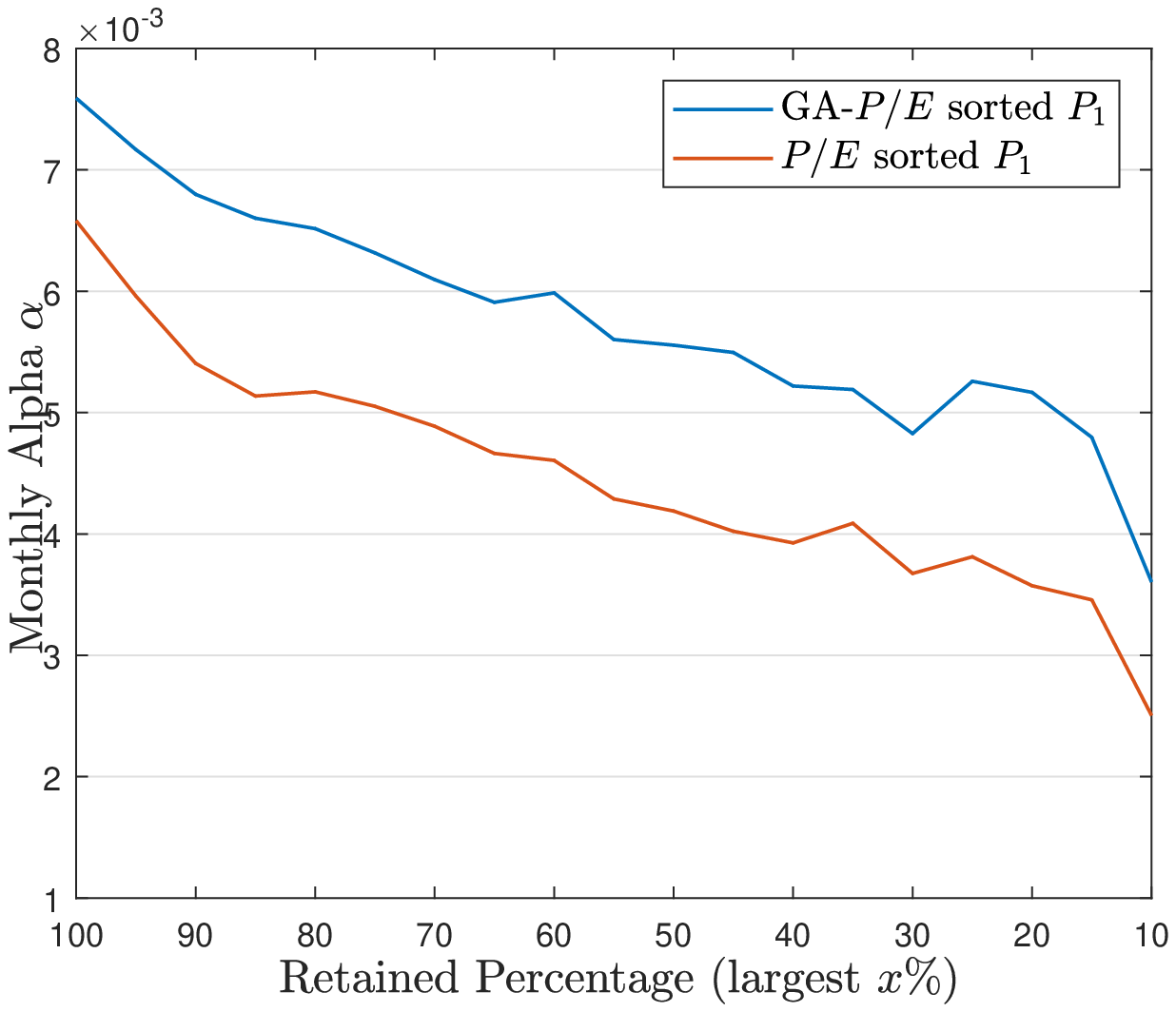}
\caption{Average annual return (top left), Sharpe ratio (top right) and monthly $\alpha$ (bottom) for the GA-$P/E$ (blue) and $P/E$ (orange) sorted $P_1$ portfolios as the stock pool is restricted to the largest $x\%$ of market capitalisations.}
\label{alpha slope}
\end{figure}

\noindent From the above charts it is apparent that the outperformance of the low GA-$P/E$ stocks over the low $P/E$ stocks cannot be attributed to a bias for smaller cap stocks. As we increasingly restrict the pool of stocks from which the portfolios are selected, the margin of outperformance, as measured by our three metrics remains fairly consistent. Indeed, if anything the gap between the two portfolios increases as we eliminate the very smallest microcap stocks from our selection pool.

\section{Conclusion}

This paper has proposed a new growth-adjusted $P/E$ ratio to improve on the noted shortcomings of the existing price-earnings based value measures.  To start, a simple dimensional analysis revealed that the standard $P/E$ ratio is not a simple dimensionless ratio, but is a period of time, that being the number of years required for a company's accumulated earnings per share to equal the  current share price, assuming earnings remain constant.

Whilst the PEG ratio provides a simple attempt to judge an appropriate $P/E$ given a certain level of growth, it is revealed as a meaningless quantity having the dimensions of time/percentage growth rate.  What is more the PEG is unable to distinguish the relative value of paying, for instance, a $P/E$ of 10 for a growth rate of 10\% or a $P/E$ of 20 for a growth rate of 20\%, in each case the PEG being 1. The GA-$P/E$ shown in \eqref{GA-$P/E$}  has the merit of remaining a period of time, as is the case with the  $P/E$ ratio, however the GA-$P/E$ adjusts for a series of growing earnings. 


The paper proceeds with an extensive anaylsis of stock market returns of the US markets over the period from 1990 to 2015. It shows that, divided into quintile portfolios, the lowest quintile GA-$P/E$ stocks generate superior returns to the higher quintiles.  This is consistent with the results for the conventional $P/E$, however the differential in returns obtained with the  GA-$P/E$ is shown to be greater than that obtained using the $P/E$. Furthermore, the return provided by the lowest quintile GA-$P/E$ portfolio exceeds that for the lowest quintile $P/E$ portfolio. 

We further regress the returns of each set of portfolios against the standard Fama-French risk factors.  Noting that each of the GA-$P/E$ portfolios may contain stocks of arbitrary $P/E$,  we find these portfolios display a uniform exposure to the traditional HML value risk factor . This contrasts with the results for the $P/E$ sorted portfolios which exhibit increasing exposure to HML as we move from high through to low $P/E$. The GA-$P/E$ portfolios as a whole exhibited greater variance in their exposures to the SMB size factor as compared to those for the set of $P/E$ portfolios. In particular, the lowest GA-$P/E$ portfolio produced a higher loading to the size factor than was observed for the comparable $P/E$ portfolio. However, this was not found sufficient to fully explain the outperformance of the low GA-$P/E$ portfolio, as evidenced by its higher alpha value. To invesitgate this size dependence further, we concluded with a comparative study of lowest GA-$P/E$ and $P/E$ quintiles, whilst sequentially restricting our stock universe to include those stocks with the largest market capitalisations. As the market capitalisation threshold was increased, we found that the lowest GA-$P/E$ quintile continued to outperform the lowest $P/E$ quintile, as measured by absolute return, risk adjusted return (Sharpe ratio) and excess return (Fama-French alpha).

Whilst we believe that this market wide analysis indicates the superiority of the GA-$P/E$ to conventional $P/E$ in forecasting returns, we believe also that the utility of the GA-$P/E$ will be particularly obvious in comparing similar stocks with differing $P/E$s and growth rates,  e.g. similar stocks in the same sector (oil majors etc) and specialised subgroups (e.g. the FAANGS etc).

\bibliographystyle{plainurl}
\bibliography{bibby3}

\end{document}